\documentclass[fleqn,usenatbib]{mnras}

\usepackage{newtxtext,newtxmath}
\usepackage[T1]{fontenc}
\usepackage{graphicx}	% Including figure files
\usepackage{amsmath}	% Advanced maths commands
\usepackage[english]{babel}
\newcommand{\myr}{$\rm mas\,yr^{-1}$}
\newcommand{\kms}{$\rm km\,s^{-1}$}
\newcommand{\kmskpc}{$\rm km\,s^{-1}\,kpc^{-1}$}

\title[Mapping the kinematic parameters of the Galaxy] {Mapping the kinematic parameters of the Galaxy from the Gaia EDR3 red giants and sub-giants}

\author[Fedorov et al.]{
P. N. Fedorov,$^{1}$\thanks{E-mail: pnfedorov@gmail.com (PNF)}
V. S. Akhmetov,$^{1,2}$\thanks{E-mail: akhmetovvs@gmail.com (VSA)}
A. B. Velichko,$^{1}$
A. M. Dmytrenko$^{1}$ \newauthor
and S. I. Denyshchenko$^{1}$
\\
$^{1}$Institute of astronomy of V.N.Karazin Kharkiv national university, Svobody sq. 4, 61022, Kharkiv, Ukraine\\
$^{2}$INAF-Osservatorio Astrofisico di Torino, Via Osservatorio 20, Pino Torinese, Turin, I-10025, Italy
}

\date{Accepted XXX. Received YYY; in original form ZZZ}

\pubyear{2022}

\begin{document}
\label{firstpage}
\pagerange{\pageref{firstpage}--\pageref{lastpage}}
\maketitle

\begin{abstract}
We present the results of a kinematic analysis of red giants and subgiants whose centroids are in the plane of our Galaxy. For this, the positions, parallaxes, proper motions, and radial velocities of these stars from the $Gaia$ EDR3 catalog were used. We applied two approaches to obtain kinematic parameters. The first approach -- solving the equations of the Ogorodnikov--Milne model with respect to 12 kinematic parameters -- is generally accepted, but has a number of disadvantages. The second approach developed by us is to find the components of galactocentric centroid's velocity and their partial derivatives with respect to coordinates directly from differential equations for the stellar velocity field. To calculate the kinematic parameters by the methods mentioned above, same stellar samples were used. From these samples spherical regions with a radius of 1 kpc were selected, the centers of which were located strictly in the Galactic mid-plane at the nodes of the coordinate grid ($x_{gal}$, $y_{gal}$) of a rectangular Galactocentric coordinate system with step 100 pc.
The region of the Galaxy under study occupies the coordinate interval $115^\circ < \theta < 245^\circ$, 0 kpc$ < R < $16 kpc, -1 kpc$ < z < $1 kpc. We show the behavior of local kinematic parameters as well as global parameters such as the circular velocity of stars as a function of galactocentric coordinates. For the first time, the components of centroids' spatial velocities and all their partial derivatives as well as their behavior as a function of galactic coordinates have been derived.
The behavior of the $\partial V_R/\partial \theta$ and $\partial V_\theta/\partial \theta$ parameters as a function of galactic coordinates has been derived for the first time.

\end{abstract}

\section{Introduction}
\label{sec:intro}

The third release of the $Gaia$ mission, $Gaia$~EDR3 catalogue (\cite{Prusti2016, Brown2020}), made new data available to study the kinematics of stars that belong to the Milky Way. The presence of parallaxes, radial velocities, and proper motions of stars in this release make it possible to obtain new information about the stellar kinematics in our Galaxy. The efficiency of the analysis of the stellar velocity field essentially depends on the quality and quantity of the observational material. Obviously, $Gaia$ EDR3 has an unprecedented accuracy of its astrometric parameters, about three orders of magnitude higher than in the Hipparcos project. The number of observed objects in these projects are not comparable.

Radial velocities and parallaxes are especially valuable data for kinematic studies, and, together with the proper motions of stars, make it possible to analyze the three--dimensional velocity field. This entails the emergence of opportunities to determine some global Galactic kinematic parameters, such as the form of rotation curve, the distance to the center of the Galaxy, and the determination of the kinematic centers of rotation of various stellar systems. Although there are certain difficulties in interpreting the results obtained that caused, for example, by the discrepancy between the distances to sources determined from the $Gaia$ parallaxes and from the Bayesian method (for instance, \cite{Bailer-Jones2021}), usage of different estimates of the distance R$_\odot$ to the Galactic center, or an insufficient number of stars with known radial velocities ( about 7.21 million), the relevance of these works is beyond doubt.

In the paper by \cite{Fedorov2021}, we presented the results of determining, within the Ogorodnikov--Milne (O–-M) model (\cite{Ogorodnikov1965, Milne1935, Clube1972, duMont1977, Miyamoto1993, Miyamoto1998}), the kinematic parameters of stellar systems whose centroids are located only along the direction of the Galactic center - the Sun - the Galactic anticenter. 

In this paper, we present the results of deriving the kinematic parameters in that part of the Galactic plane, which is provided with six astrometric parameters from $Gaia$ EDR3. We derive the kinematic parameters from an analysis of the kinematics of red giants, whose velocities are considered relative to a fixed center. To do this, knowing the velocities of stars relative to the barycenter of the Solar System, as well as the velocity of the Sun $V_\odot$ relative to the Galactic center, we exclude the latter from the relative velocities of all stars under study, and thus, set the field of velocities of stars in a fixed rectangular coordinate system.

For calculations, the following values of the Solar velocity relative to the Galactic center were used: $(V_{x,\odot}, V_{y,\odot}, V_{z,\odot}) = (11.1,243.13,8.31) $\kms, assuming that $R_\odot$ = 8.0 kpc (\cite{Vallee2017}). The $V_{y,\odot}, V_{z,\odot}$ components were determined by \cite{Reid2020} from the proper motion of the Sagittarius-A source. The $V_{x,\odot}$ component was taken from the paper by \cite{Schonrich2010} devoted to determination of the velocity of the local standard of rest.

In Section \ref{sec:preparation}, we present the basic formulas that are used to transform the spatial motions of stars from the local spherical coordinate system with its origin in the Solar barycenter to a local rectangular one with its origin located at some greed node. Section \ref{sec:methods} presents the Ogorodnikov--Milne formulas in a rectangular coordinate system and their connection with global Galactic kinematic parameters. Formulas for calculating the average values of Galactocentric velocities and their gradients are also presented.
Section \ref{sec:analysis} contains brief analysis and comparison of kinematic parameters obtained by two methods.

\section{The data preparation. The scheme of constructing the grid of nodes.}
\label{sec:preparation}

To derive the kinematic parameter the same stellar sample of red giants and sub--giants from $Gaia$~EDR3 as in the work by \cite{Fedorov2021} was used. The total amount of stars under study is $\sim$4.5 million.

Using the formula below, the velocity vector components were transformed from the spherical coordinate system $(r, l, b)$ to the rectangular Cartesian one $(x, y, z)$.
\begin{align}
  & V_x = V_r\,{\rm cos}\,l\,{\rm cos}\,b - V_l\,{\rm sin}\,l - V_b\,{\rm cos}\,l\,{\rm sin}\,b \nonumber \\
  & V_y = V_r\,{\rm sin}\,l\,{\rm cos}\,b + V_l\,{\rm cos}\,l - V_b\,{\rm sin}\,l\,{\rm sin}\,b  \\
  & V_z = V_r\,{\rm sin}\,b + V_b\,{\rm cos}\,b, \nonumber
\end{align}
where $V_r$ is the radial velocity, $V_l = k\,r\,\mu_l\,{\rm cos}b, V_b = k\,r\,\mu_b\,$ are components of stellar proper motion, $k$ = 4.74057 is the transformation factor from \myr to \kmskpc. The distance $r$ is calculated from the $Gaia$~EDR3 parallax as $1/\pi$.

\begin{figure*}
\includegraphics[width = 88mm]{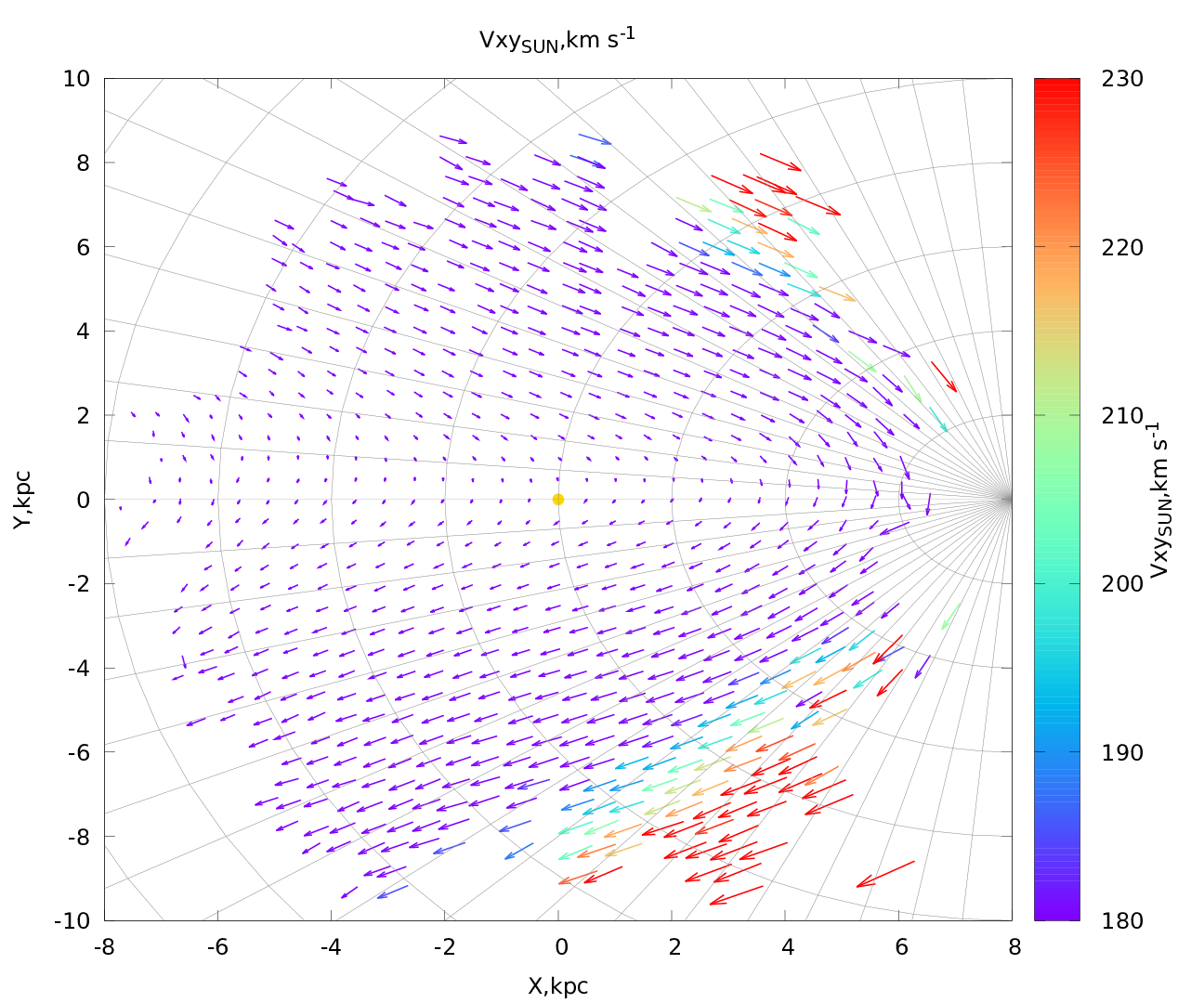}
\includegraphics[width = 88mm]{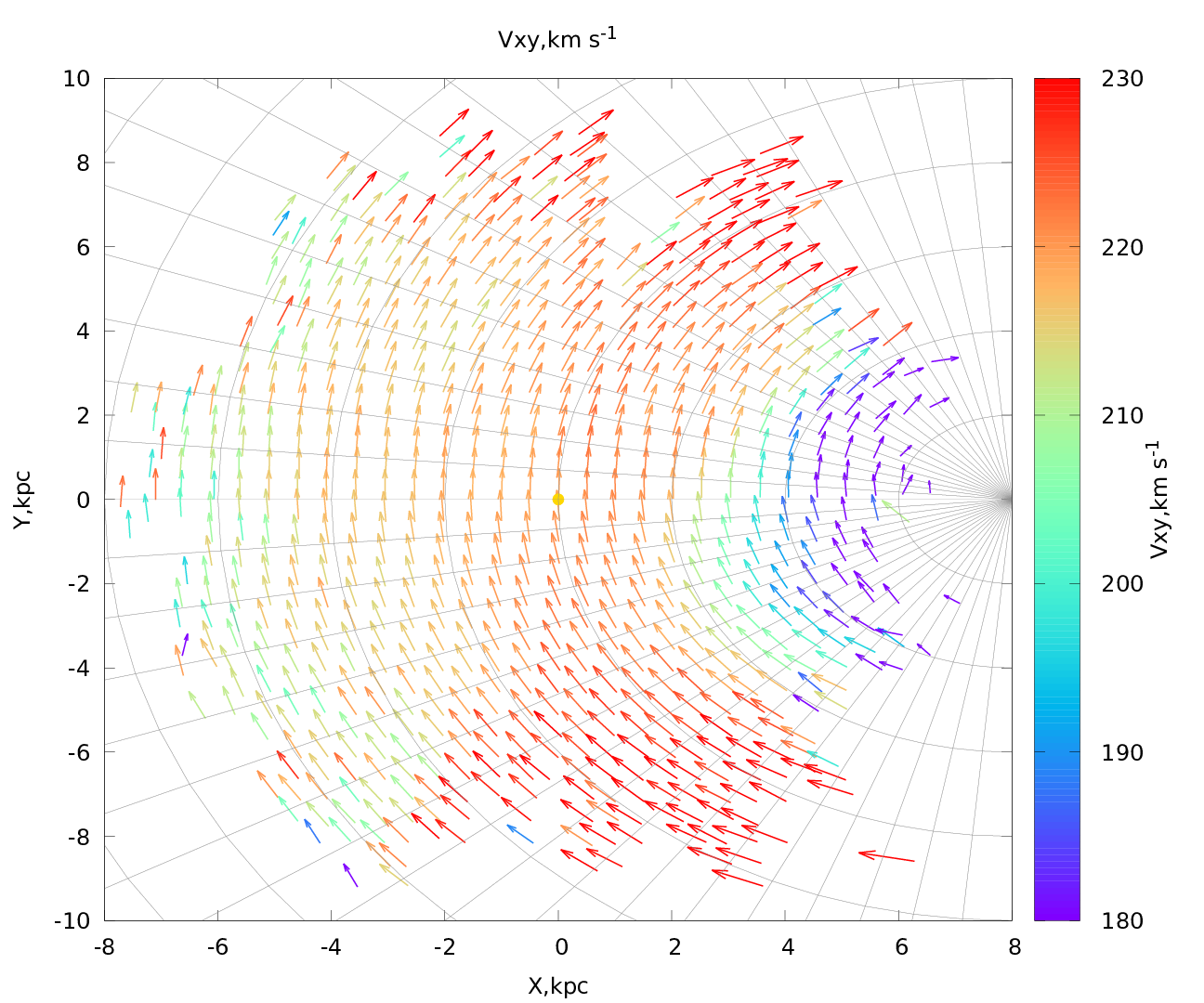}
\caption{Vector diagrams of the centroids velocity components observed (left panel) and in motionless coordinate system (right panel) as a function of the rectangular Galactic and cylindrical Galactocentric coordinates. The module of the velocity vector was scaled for clear visualization. $R_{\sun}$ is taken to be 8.0~kpc and showed yellow point.}
\label{fig:Vxy}
\end{figure*}

These formulas represent the components of the relative velocity vector of the Sun and a specific star in a rectangular coordinate system, the origin of which is located at the barycenter of the Solar System and determine the relative velocity field of the stars (Fig. \ref{fig:Vxy}, left panel).

Taking the values of the components of the solar velocity vector $V_{x,\odot}, V_{y,\odot}, V_{z,\odot}$ relative to the Galactic center known from (\cite{Reid2020, Schonrich2010}), we calculate the velocities of individual stars that determine velocity field relative to the motionless coordinate system:

\begin{align}
    & V_{x,gal} = V_{x} + V_{x,\odot}, \nonumber\\
    & V_{y,gal} = V_{y} + V_{y,\odot},          \\
    & V_{z,gal} = V_{z} + V_{z,\odot}  \nonumber
\end{align}

Further, we analyze this velocity field (Fig. ~\ref{fig:Vxy} right panel).
Two approaches to the analysis are applied: the main one, using the velocities of individual stars, and the additional one to control the correctness of the results obtained in the main approach, using the averaged velocities of stars. With an additional approach in the Cartesian coordinate system, the space containing the stars of our sample was divided into cubic cells 100×100×100 pc in size. The astrometric parameters of each cell (fictitious star), i.e. the positions $ (x, y, z)$ and spatial velocities ($V_x, V_y, V_z$ ) were obtained by averaging the individual astrometric parameters of the stars that get into the cell.

As was noted in \cite{Fedorov2021}, a rectangular Galactic coordinate system can be introduced at any arbitrary point of the Galactic plane, provided that the spatial coordinates and components of the spatial velocity are known for this point and for the stars located in its vicinity. The transition from the Cartesian coordinate system with the origin at the barycenter of the Solar System to the local Cartesian coordinate system with the origin at the chosen node also requires knowledge of the distance from the center of the Galaxy to the Sun. In this work, we take the value $R_\odot$ = 8.0 kpc (\cite{Vallee2017}) to be able to compare our results with our previous work.

\begin{figure}
   \centering
\resizebox{\hsize}{!}
   {\includegraphics{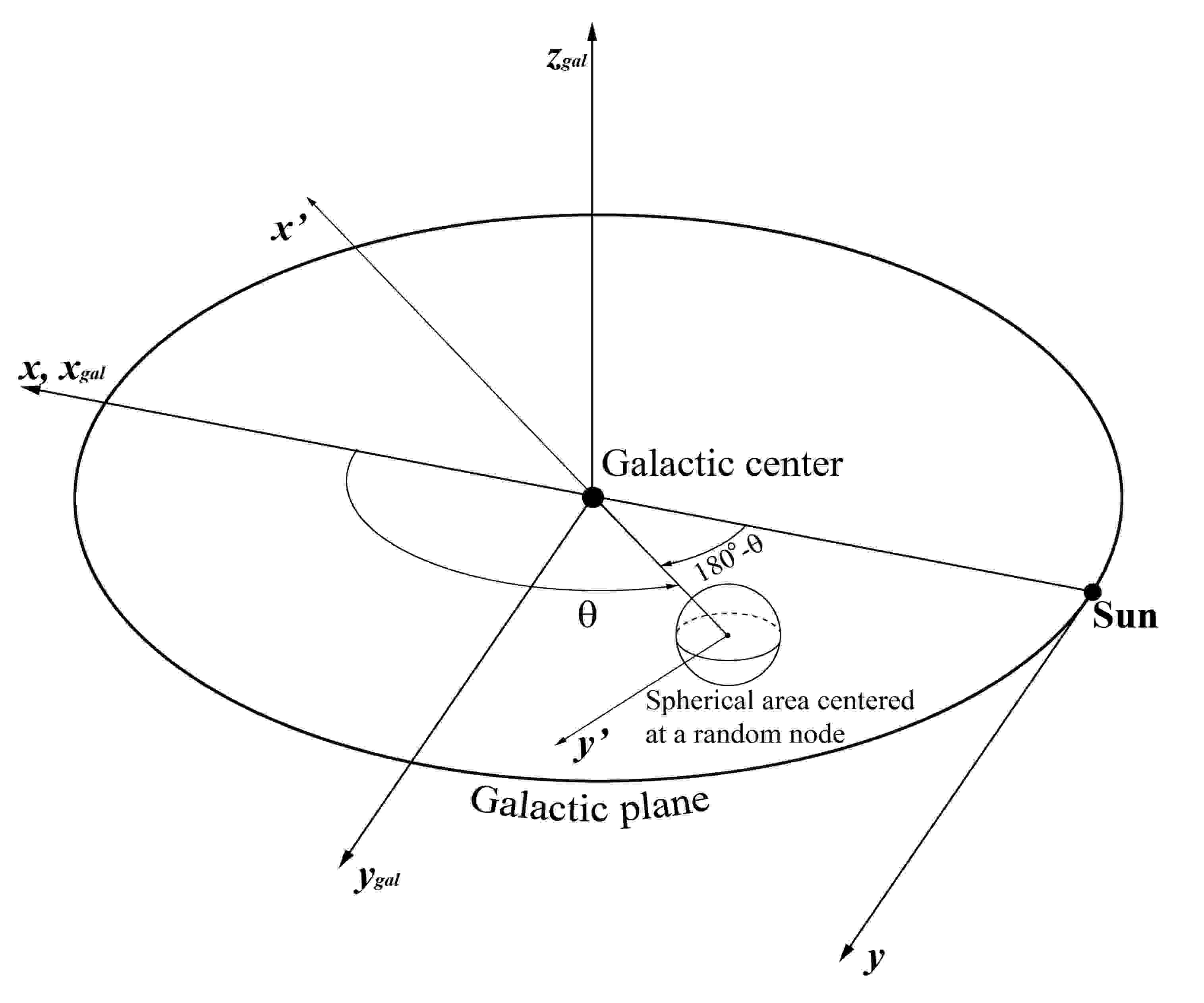}}
  \caption{The scheme of coordinate systems used. $x, y, z$ are coordinates of the local rectangular Galactic coordinate system with its origin in the Solar system's barycenter. $x_{\rm gal}, y_{\rm gal}, z_{\rm gal}$ are coordinates of the rectangular Galactic coordinate system with the origin in the Galactic center. The spherical region surrounding a specified node and segregating a stellar system, is shown. All kinematic parameters for the stellar system are derived in the coordinate system $x', y', z'$, where $x'$-axis is directed to the Galactic center, $y'$ is always coincides with the direction of the Galactic rotation at the point of the Galactic plane. All such nodes are located strictly in the Galactic mid-plane.}
\label{fig:mw-scheme}
\end{figure}

In the local Cartesian coordinate system (Fig. \ref{fig:mw-scheme}), the $x'$ axis is always directed from a specific point to the center of the Galaxy, the $y'$ axis is in the direction of Galactic rotation and is perpendicular to $x'$ , and the $z'$ axis is always perpendicular to the Galactic plane. This requirement is mandatory, since only in this case the mechanical meaning of the kinematic parameters of the O--M model remains the same as if the origin of the coordinate system would locate in the Sun. This approach makes it possible to compare and analyze the kinematic parameters of centroids at different angles $\theta$. Here it is appropriate to note that the unit vectors of the local Cartesian coordinate system actually coincides with that of the cylindrical coordinate system originating at the center of the Galaxy, but the unit vectors ${\bf i} = -{\rm \bf e}_R$ and ${\bf j} = -{\rm \bf e}_\theta$ are oppositely oriented. The transition from the Cartesian coordinate system with the origin in the Sun to the local Cartesian system with the origin in the centroid corresponds to the displacement of the fictitious observer from the barycenter of the Solar System to the point specified by the origin of the chosen coordinate system.

\section{Methods for determination of kinematic parameters}
\label{sec:methods}

First of all, we single out spherical regions with a radius of 1.0 kpc, whose centers we place at the nodes of a rectangular grid specified in the galactic plane. The grid nodes are separated from each other along the coordinates $x$ and $y$ by distance of 100 pc. Each sphere circumscribed around a given node contains stars that are located at distances not exceeding 1 kpc from the node. After that, we compute the components of the centroids' average velocities and the partial derivatives of these Components with respect to the coordinates (this is what the kinematic parameters are) both in the local rectangular ($V_x, V_y, V_z$) and in the Galactocentric cylindrical ($V_R, V_\theta, V_Z$) coordinate systems.

It should be noted that the spatial velocities of the centroids can be distorted by the inaccuracy of the accepted value of the Solar galactocentric velocity vector. However, certain partial derivatives of velocities with respect to the coordinates do not depend on this value, since the inaccuracies of the solar velocity components enter as constants into the velocity components of all stars under study. Thus, the figures below demonstrate the behavior of the kinematic parameters depending on the galactic coordinates, which is not distorted by the inaccuracy of the Galactocentric velocity of the Sun. In this case, the zero point (constant shift) of these parameters obviously depends on the accepted value of the solar Galactocentric velocity vector.

In this work, we used two approaches to determine the kinematic parameters of the Galaxy. In the first approach, the equations to determine the parameters of the O--M model were used. They are written in a local Cartesian coordinate system with the origin at the selected node $x', y', z'$.

\begin{align}
& V_x = M^+_{11}x' + M^+_{12}y' - \omega_3y' + M^+_{13}z' + \omega_2z' - U_x \nonumber \\
& V_y = M^+_{12}x' + \omega_3x' + M^+_{22}y' + M^+_{23}z' - \omega_1z' - U_y \\
& V_z = M^+_{13}x' - \omega_2x' + M^+_{23}y' + \omega_1y' + M^+_{33}z' - U_z \nonumber 
\end{align}

For the orientation of a specific local coordinate system with the origin at an arbitrary point of the Galactic plane, using the coordinates of the selected point $x, y$ as well as the distance to the Galactic center $R_\odot$, we determined the angle $\phi = 180^\circ - \theta$, by which it is required to rotate the $X$ axis of the original system associated with the Sun, where the angle $\theta$ is the coordinate of the Galactocentric cylindrical system. The orientation of the so defined local coordinate system will correspond to the following conditions: the $x'$ axis is directed to the center of the Galaxy, the $y'$ axis is directed along the direction of the Galactic rotation, and the $z'$ axis is perpendicular to the Galactic plane. In this case, the origins of all local coordinate systems will move relative to the fixed center, and the quantities $U_x, U_y, U_z$ will be the components of the spatial velocity of the centroids in the Cartesian Galactocentric coordinate system $x_{\rm gal}, y_{\rm gal}, z_{\rm gal}$.

In this case, the mechanical meaning of the kinematic parameters obtained in different local coordinate systems remains unchanged. The diagonal components of the symmetric matrix -- $M^+_{11}, M^+_{22}, M^+_{33}$ characterize the rates of relative expansions - contractions of the stellar system along the axes $x', y', z'$, components $M^+_{12} = M^+_{21}$, $M^+_{23} = M^+_{32}$, $M^+_{13} = M^+ _{31}$ characterize the angular deformation rates in the planes ($x', y'$ ), ($y', z'$ ) and ($z', x'$), respectively, and the values $\omega_1, \omega_2, \omega_3$ are the projections of the instantaneous angular velocity vector $\bf{\omega}$ of the stellar system under consideration onto the galactic axes $x', y', z'$. The O--M equations were solved with respect to that 12 unknowns.

The relations between the kinematic parameters of the O--M model and the partial derivatives of the centroid velocity components with respect to the Galactocentric coordinates $R, \theta, z$ is established by the following formulas:

\begin{align}
& M^+_{11} = \partial V_R/\partial R                                  \\
& (\omega_3 + M^+_{12}) = \partial V_\theta/\partial R = -\partial V_{\rm rot}/\partial R \label{eq:dVtheta_dR_om}\\
& (\omega_2 - M^+_{13}) = \partial V_z/\partial R    \label{eq:dVz_dR_om}                 \\
& (\omega_3 - M^+_{12}) = V_\theta/R - 1/R \partial V_R/\partial\theta \label{O3-M12p} \\
& M^+_{22} = V_R/R + 1/R \partial V_\theta/\partial\theta = V_R/R - 1/R \partial V_{\rm rot}/\partial\theta \\
& (\omega_1 + M^+_{23}) = -1/R \partial V_z/\partial\theta  \label{warp}\\
& (\omega_2 + M^+_{13}) = -\partial V_R/\partial z  \label{eq:dVR_dz_om}\\
& (\omega_1 - M^+_{23}) = \partial V_\theta/\partial z = -\partial V_{\rm rot}/\partial z \label{eq:dVtheta_dz_om}\\
& M^+_{33} = \partial V_z/\partial z  \label{eq:dVz_dz}
\end{align}

Obviously, these formulas do not allow one to determine the partial derivatives of the radial and azimuth velocities from the coordinate angle $\theta$ from combinations of the O--M model parameters. Therefore, some more or less justified assumptions have to be introduced regarding these derivatives. So, in particular, from the combination of $\omega_3$ and $M^+_{12}$, derived within the O--M model, we get:

\begin{equation}
    V_\theta = (\omega_3 - M^+_{12})R + \partial V_R/\partial\theta
\end{equation}

Since $V_\theta$ is directed opposite to the circular rotation velocity of the Galaxy $V_{\rm rot}$, we can write

 \begin{equation}
     V_{\rm rot} = -V_\theta = (M^+_{12} - \omega_3)R - \partial V_R/\partial\theta
 \end{equation}

If we assume, as is done in the Oort--Linblad model, that $\partial V_R/\partial\theta = 0$, i.e., the star system under consideration is axisymmetric, we obtain a frequently used formula for determining $V_{\rm rot}$  (\cite{Ogorodnikov1965, Milne1935, Clube1972, duMont1977, Miyamoto1993, Miyamoto1998, Vityazev2005, Vityazev2009, Bobylev2011}), through the Oort constants A and B:

\begin{equation}
    V_{\rm rot} = (A-B)\,R
    \label{eq:Vrot}
\end{equation}

As already shown in \cite{Fedorov2021}, the behavior of the curve of centroids' circular velocity determined from the formula \ref{eq:Vrot} and by using the value of the Solar velocity relative to the Galactic center \cite{Reid2020}, differ markedly. The reason for this difference, apparently, is the assumption that the stellar velocity field is axisymmetric.

The second approach is to find the Galactocentric components of the average centroid velocities and their partial derivatives with respect to coordinates from the equations:

\begin{align}
\label{eq:derivatives}
& V_R (R,\theta, z) = \overline{V_R} +\frac{\partial V_R}{\partial R}  R + \frac{\partial V_R}{\partial \theta}  \theta + \frac{\partial V_R}{\partial z}  z;\nonumber \\
& V_\theta (R,\theta, z) = \overline{V_\theta} +\frac{\partial V_\theta}{\partial R}  R + \frac{\partial V_\theta}{\partial \theta}  \theta + \frac{\partial V_\theta}{\partial z} z; \\
& V_z (R,\theta, z) = \overline{V_z} +\frac{\partial V_z}{\partial R} R + \frac{\partial V_z}{\partial \theta} \theta + \frac{\partial V_z}{\partial z} z;\nonumber
\end{align}

This approach makes it possible to obtain estimations of all kinematic parameters, including those that are fundamentally impossible to determine from the O--M model. These parameters include the gradients of the radial $\frac{\partial V_R}{\partial \theta}$ and the azimuthal  $\frac{\partial V_\theta}{\partial \theta}$ velocity along the coordinate angle $\theta$.

\begin{figure*}
\includegraphics[width = 88mm]{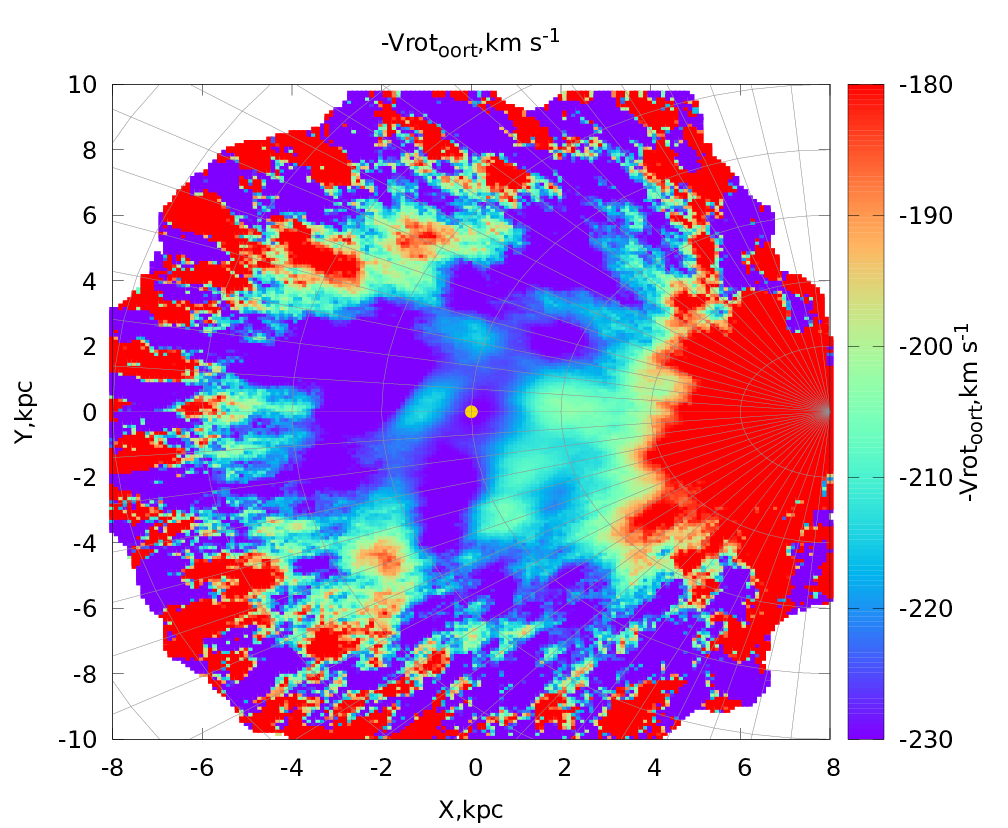}
\includegraphics[width = 88mm]{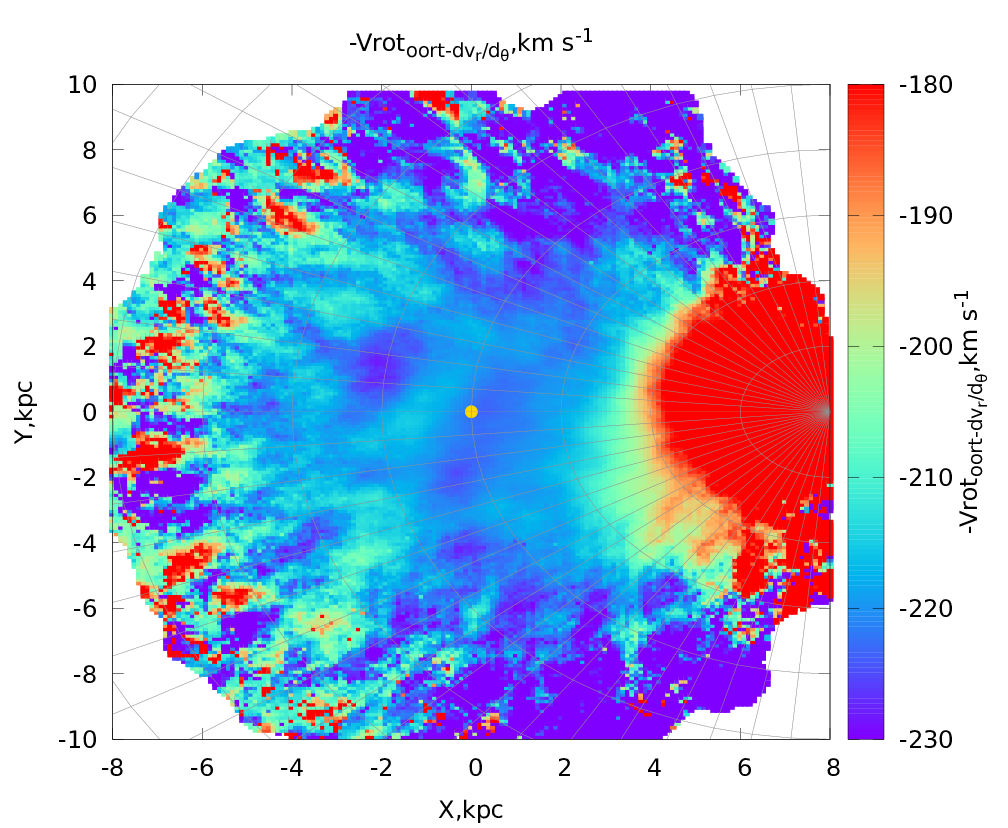}
\caption{Rotation component of the centroids velocity $V_{\rm rot}$ with opposite sign, determined from the Oort constants (equation \ref{eq:Vrot}) on the assumption that $\partial V_R / \partial \theta$ is equal to 0 (left panel) and taking into account $\partial V_R / \partial \theta$ computed from equations \ref{eq:derivatives} (right panel) as a function of the Galactic coordinates. $R_{\sun}$ is taken to be 8.0~kpc and indicated by the yellow circle.}
\label{fig:Vrot}
\end{figure*}

To check the assumption made about axisymmetry, we presented in Fig. ~\ref{fig:Vrot} the velocity $V_{\rm rot}$ = $-V_\theta$, determined from the Oort constants under the assumption $\partial V_R/\partial\theta = 0$  (left panel ) and defined by the formula \ref{O3-M12p} using $\partial V_R/\partial\theta$ Fig. ~\ref{fig:Vrot} (right panel) determined from equations \ref{eq:derivatives}.

\begin{figure}
\centering
\resizebox{\hsize}{!}
{\includegraphics{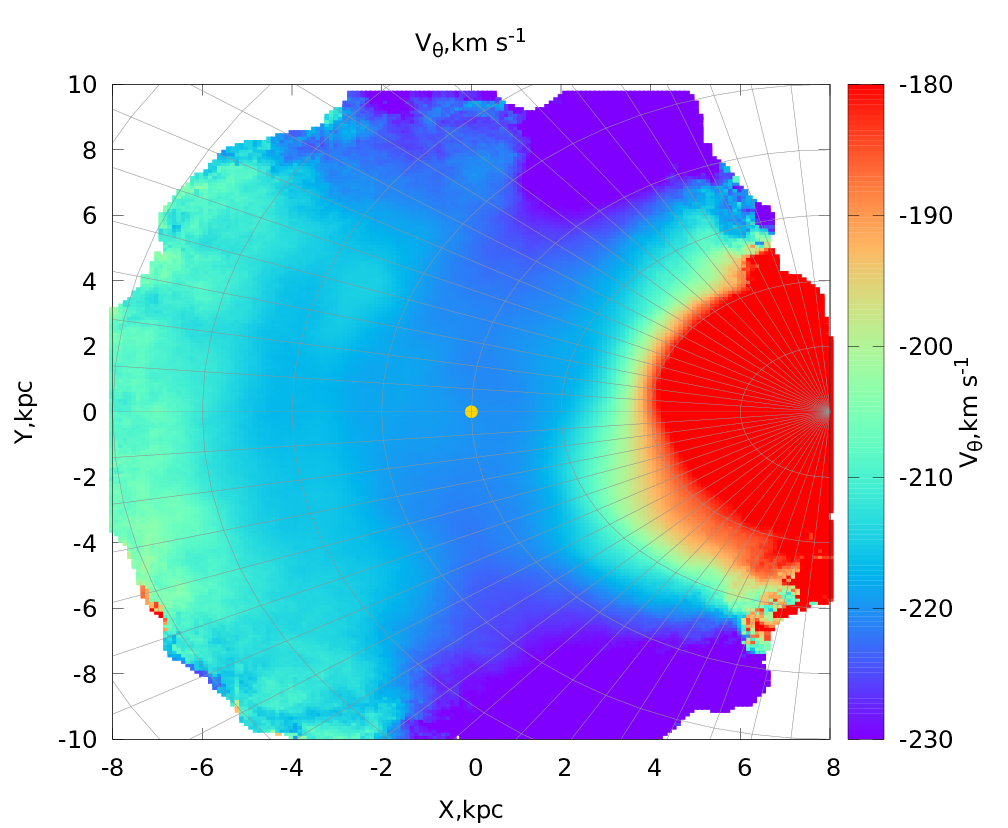}}
\caption{Rotation components of the centroids velocity $V_\theta$ determined from equations \ref{eq:derivatives} as a function of the Galactic coordinates. $R_{\sun}$ is taken to be 8.0~kpc and showed yellow point.}
\label{fig:Vtheta}
\end{figure}

It can be clearly seen from the Fig. \ref{fig:Vrot} that $V_{\rm rot}$ determined from the Oort constants and corrected for $\partial V_R/\partial\theta$ (right panel) noticeably approaches the values of $V_{\rm rot}$ determined from the spatial velocities of centroids from equations \ref{eq:derivatives} (Fig. ~\ref{fig:Vtheta}) . It is quite obvious that the circular velocity of centroids $V_{\rm rot}$ is determined most accurately when using spatial centroid velocities, since in this case we do not use any assumptions and do not impose any restrictions on the velocity field.

\section{Analysis of the results derived}
\label{sec:analysis}

The kinematic parameters derived in the work are presented for clarity in the form of color maps. This way of visualizing kinematic parameters, the amplitude of which is represented by color, is essentially illustrative.

On the color maps, the coordinate plane $x, y$ coincides with the Galactic plane. The kinematic parameters related to a particular point $x, y$ of this plane are obtained in the local rectangular coordinate system, the $x', y'$ axes, which coincide with the $R$ and $\theta = 180^\circ - \phi$ axes of the cylindrical coordinate system with origin at the center of the Galaxy, but opposite to them in direction.

\subsection{Radial velocity of centroids $V_R$}

\begin{figure}
\centering
\resizebox{\hsize}{!}
   {\includegraphics{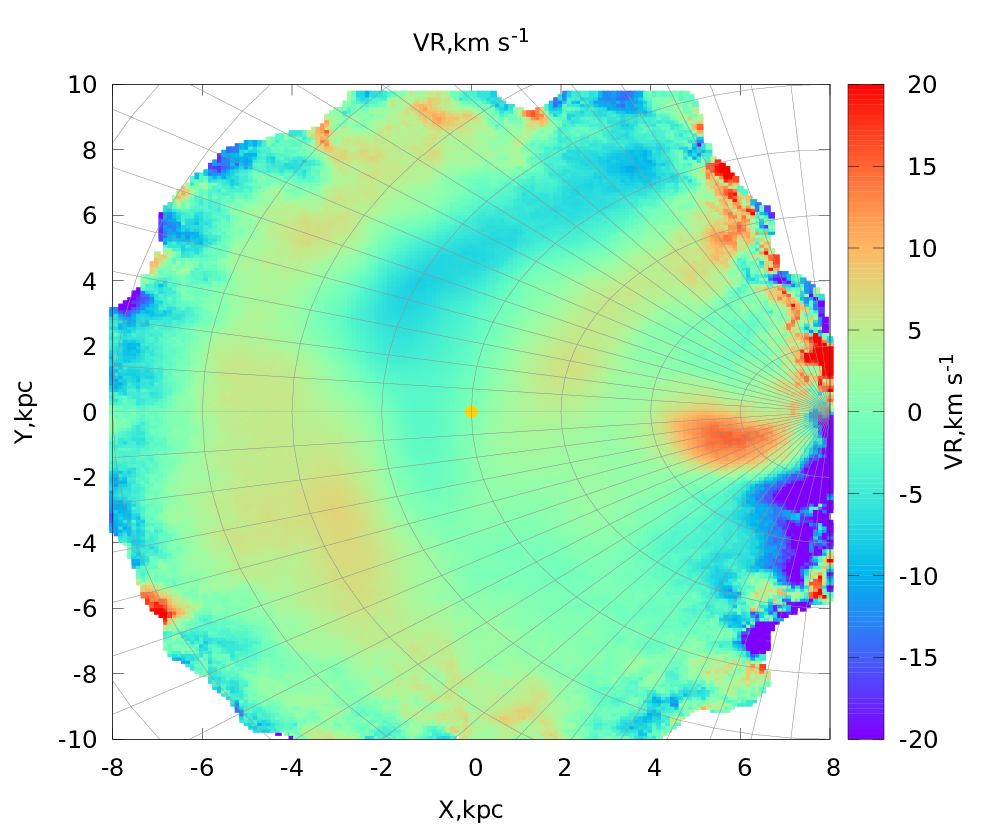}}
  \caption{Radial component of the centroids velocity as a function of the Galactic coordinates. $R_{\sun}$ is taken to be 8.0~kpc and showed yellow point.}
\label{fig:VR}
\end{figure}

In Fig. ~\ref{fig:VR} the radial component of the centroids' velocity as a function of the galactic coordinates is shown. As can be seen from the figure, this component in the presented part of the galactic plane has velocities, the values of which vary from about -10 to +10 \kms. Structural features with increased or decreased velocities are observed, arranged in the form of concentric circles. Although the velocities in these areas are relatively small, nevertheless, it is visually obvious that they keep certain patterns. It is likely that they are a consequence of global kinematic regularities. The radial velocity of the centroid of the Sun, as can be seen from the figure, is close to zero.

The influence of the contribution of non-zero values of this velocity component to the total velocity of centroids in the Galactic plane practically absent, which can be seen in vector diagram Fig. \ref{fig:Vxy} (right panel). Due to the absence of this influence, the velocity vectors towards the Galactic center are practically perpendicular throughout the Galactocentric distances from approximately 2 to 16 kpc.

\subsection{Circular velocity of centroids $V_\theta$}

The change in the velocity component $V_{\rm rot}=-V_\theta$ along the direction of the Galactic center  -- the Sun is the rotation curve of the Galaxy in this direction. Fig. \ref{fig:Vtheta} shows the rotation curves depending on the coordinate angle $\theta$. As can be seen from the figure, the direction of the center of the Galaxy -- the Sun -- the anticenter can be considered as an axis of symmetry with respect to the velocity distribution $V_\theta$ in the presented part of the galactic plane. Average values of velocities of centroids located at the same distance in the range of coordinate angles $\pm$30--40 degrees differ slightly. Outside the above range, their dependence on the angle $\theta$ is obvious. This is also clearly seen in Fig.\ref{fig:Vxy} (right panel), where the velocity of the centroids is represented as vectors, and the modulus of their velocity is indicated by color. Obviously, in this range the velocity $V_\theta$ is not only a function of the galactocentric distance $R$, but also depends on the angle $\theta$.

 At the same time, it is clearly seen that the distribution of the velocity component $V_\theta$ in the galactic plane at distances greater than 3 kpc from the center of the Galaxy and in the range of coordinate angles $\pm$30-40 degrees is relatively smooth and does not have apparent features. Numerical values of the rotation curve along the direction of the Galactic center -- the Sun practically coincides with the numerical values given in the previous article \cite{Fedorov2021}.

\begin{figure*}
 {\includegraphics [width = 88mm] {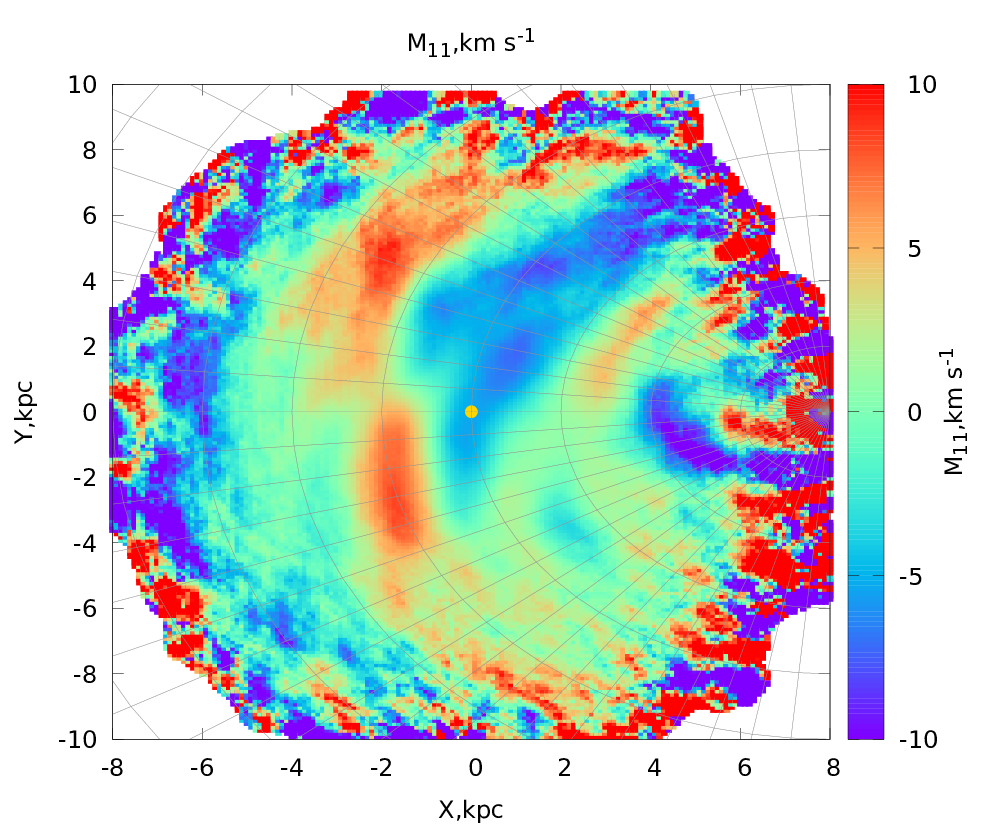}
   \includegraphics [width = 88mm] {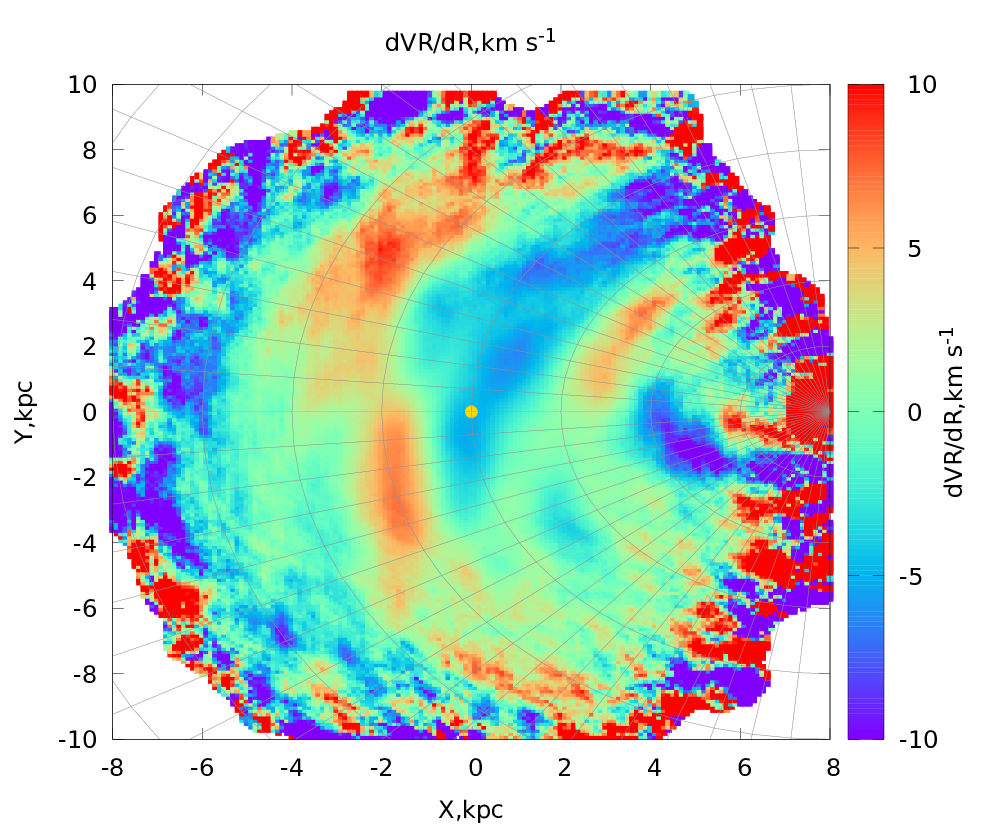}}
  \caption{Radial gradient of the radial centroids velocity component as a function of the Galactic coordinates. Left panel - $M^+_{11}$ determined from the O--M model and right panel $\frac{\partial V_R} {\partial R}$ determined from equations \ref{eq:derivatives}. $R_{\sun}$ is taken to be 8.0~kpc and showed yellow point.}
\label{fig:dVR_dR}
\end{figure*}

\subsection{Radial gradient of the Galactic expansion velocity $\partial V_R / \partial R$}

This parameter is shown in Fig. \ref{fig:dVR_dR}. On the left is the parameter obtained within the O--M model, and on the right -- from equations \ref{eq:derivatives}. As can be seen from the figure, the value and behavior of this kinematic parameter are in good agreement with each other, as well as with the values obtained in our previous work \cite{Fedorov2021} along the direction the Galactic center--the Sun -- anticenter, which indicates the correctness of our calculation methods.

It is known that, within the O--M model, $\frac{\partial V_R} {\partial R}$ is interpreted as a compression (expansion) of the stellar system along the $X$ axis ($M^+_{11}$). It can be seen from the presented Fig. \ref{fig:dVR_dR} that the distribution of this parameter in the galactic plane is has certain features in the form of extended areas with increased and decreased velocities (within $\pm$10 \kms), for the interpretation of which additional research is needed.

\subsection{Azimuth gradient of the radial velocity component $\partial V_R / \partial \theta$}

\begin{figure}
   \centering
\resizebox{\hsize}{!}
   {\includegraphics{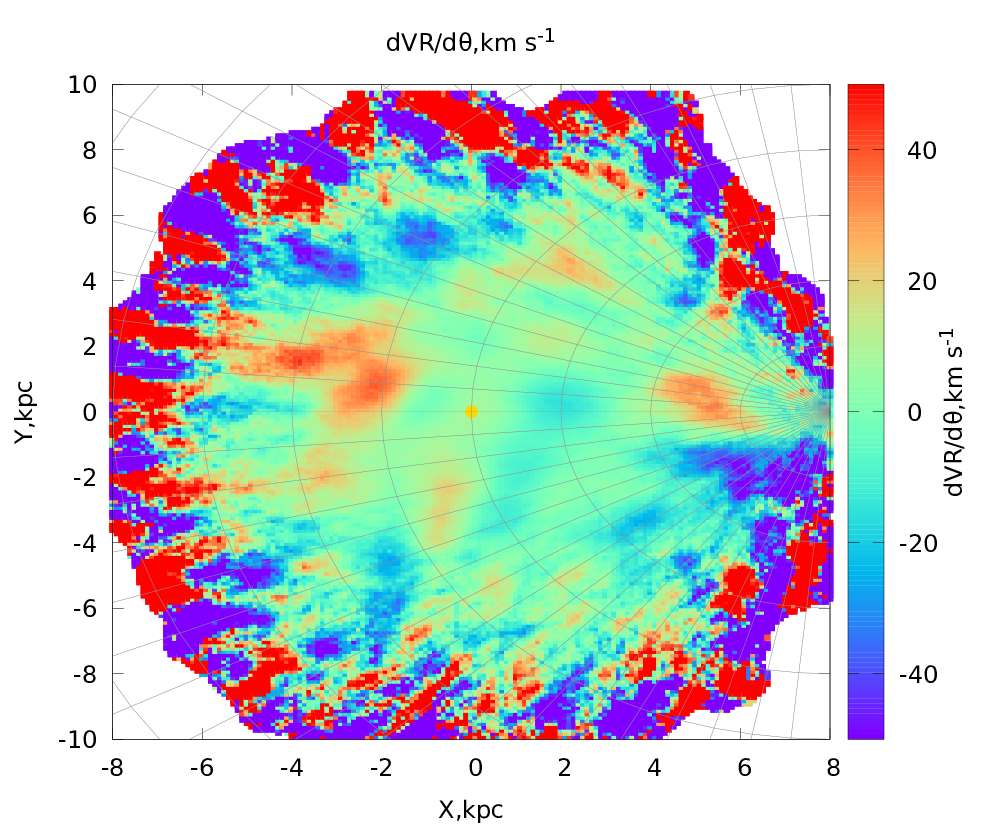}}
  \caption{Azimuth gradient of the radial centroids velocity component as a function of the Galactic coordinates. $R_{\odot}$ is taken to be 8.0~kpc and showed yellow point.}
\label{fig:dVR_dtheta}
\end{figure}

As mentioned above, this parameter cannot in principle be determined from a combination of the parameters of the O--M model. However, its definition is critical, since the value of the circular rotation velocity, determined through the parameters $\omega_3$ and $M^+_{12}$, depends on the knowledge of its value. As can be seen from Fig. ~\ref{fig:dVR_dtheta}, the assumptions about the axisymmetric galactic rotation do not correspond to reality. Even in the central regions of the considered part of the Galaxy, the gradient values differ significantly and reach values of $\pm$25--30 \kms. This, in turn, should result in significant differences in the determination of the galactic rotation velocity. This difference is shown in the left part of Fig. \ref{fig:Vrot}, where the value of $\frac{\partial V_R} {\partial \theta}$ is assumed to be 0, and in the right part of Fig. \ref{fig:Vrot}, where the value of this parameter found by formulas \ref{eq:derivatives} is taken into account.

\subsection{Vertical gradient of Galactic expansion Velocity $\partial V_R/\partial z$}

\begin{figure*}
   \centering
\resizebox{\hsize}{!}
   {\includegraphics [width = 88mm] {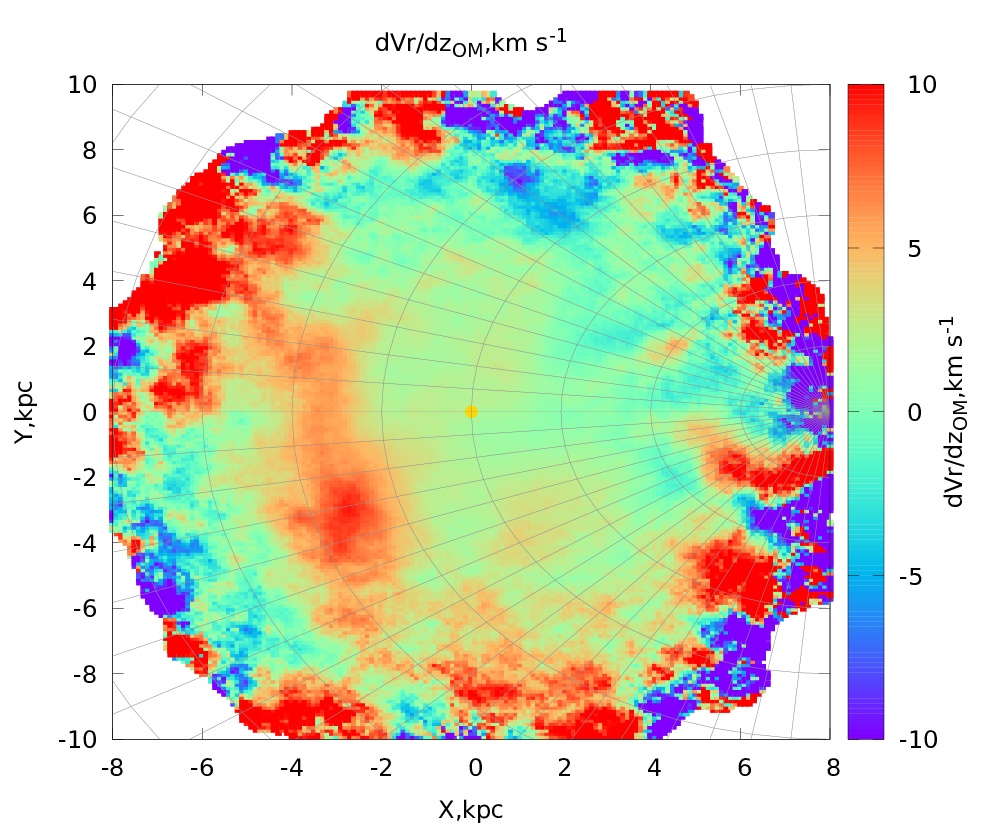}
    \includegraphics [width = 88mm] {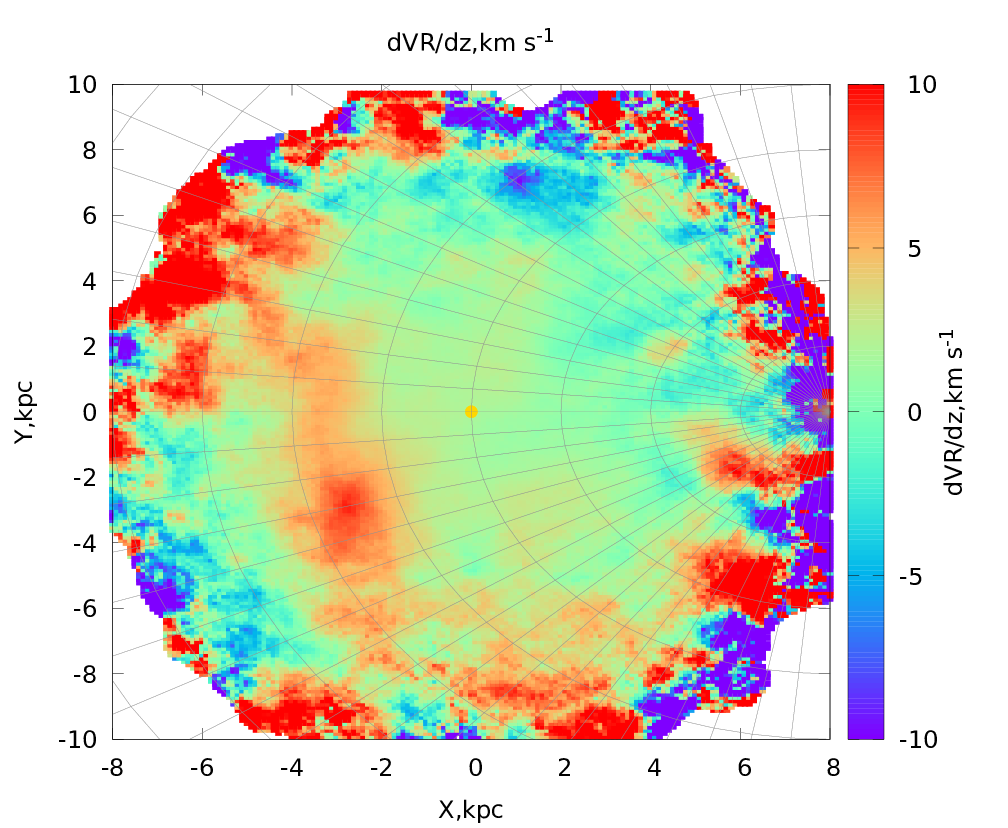}}
  \caption{Vertical gradient of the radial centroids velocity component $\frac{\partial V_R}{\partial z}$ as a function of the Galactic coordinates, derived from the O--M model parameters (left panel) and directly from the equations \ref{eq:derivatives} (right panel). $R_{\odot}$ is taken to be 8.0~kpc and showed yellow point.}
\label{fig:dVR_dz}
\end{figure*}

This parameter is shown in Fig. ~\ref{fig:dVR_dz}. Within the O--M model ( left panel), a linear combination of parameters $(\omega_2 + M^+_{13}) = -\partial V_R/\partial z \\$ represents the vertical velocity gradient of the Galactic expansion. Also, the value of this parameter can be calculated from the equations \ref{eq:derivatives} (right panel of the Fig. ~\ref{fig:dVR_dz}). As can be seen from the Fig. ~\ref{fig:dVR_dz}, the values of this parameter obtained by both methods are in good agreement with each other, and the distribution of this parameter in the galactic plane up to heliocentric distances of approximately $\pm$4~kpc is relatively smooth and has velocities within $\pm$4 \kms.

\begin{figure*}
{\includegraphics [width = 88mm] {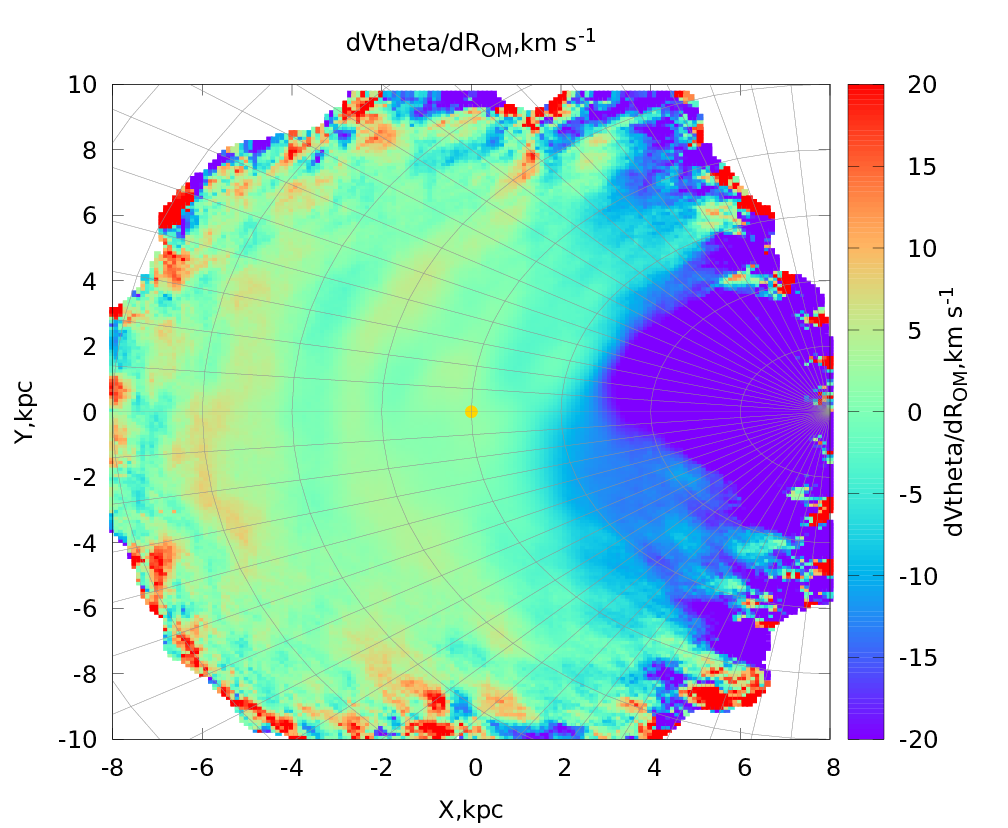}
    \includegraphics [width = 88mm] {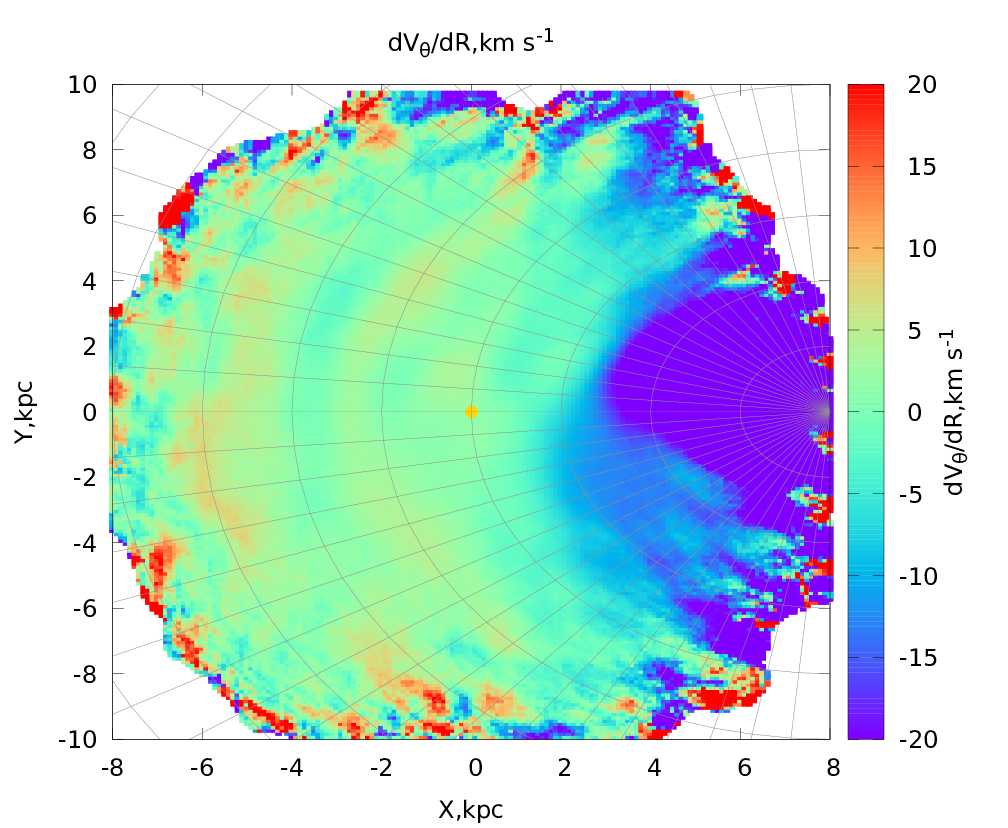}}
\caption{Radial gradient of the circular velocity derived from the O--M model parameters (left panel) and from the equation \ref{eq:derivatives} (right panel) as a function of the Galactictocentric coordinates. $R_{\odot}$ is taken to be 8.0~kpc and showed yellow point.}
\label{fig:dVtheta_dR}
\end{figure*}

 \subsection{Radial gradient of the circular velocity component $\partial V_\theta / \partial R$}

The radial gradient of the Galactic rotation velocity, or the slope of the Galactic rotation curve along the $R$ direction, is given by the equation \ref{fig:dVtheta_dR}.

The distribution of this kinematic parameter in the Galactic plane is shown in the left (O--M model, equation \ref{eq:dVtheta_dR_om}) and right (equations \ref{eq:derivatives}) panels of Fig. ~\ref{fig:dVtheta_dR}. As can be seen from the Fig. ~\ref{fig:dVtheta_dR}, the quantities $\frac{\partial V_\theta} {\partial R}$ in both cases are in good agreement with each other both in the range of Galactocentric distances $R$ from 0 to 5 kpc and within the entire range of angle $\theta$. They are nearly constant with the value approximately -15 \kms. At distances exceeding 5~kps, there is an almost abrupt change parameter values to that close to zero. The figure also clearly shows some patterns in the form of alternating ring-shaped structures with velocities from 0 to about 5-7~\kms and for the interpretation of which additional research is needed.

\subsection{Vertical gradient of the circular velocity component $\partial V_\theta / \partial z$}

\begin{figure*}
   {\includegraphics [width = 88mm] {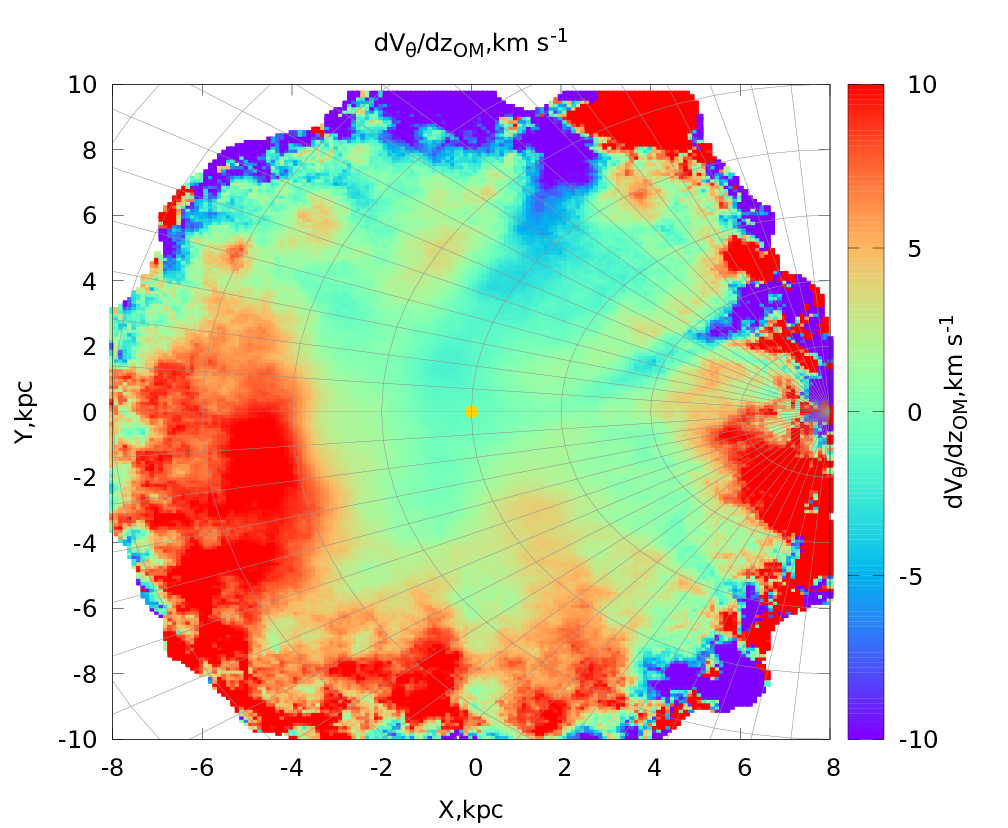}
   \includegraphics [width = 88mm] {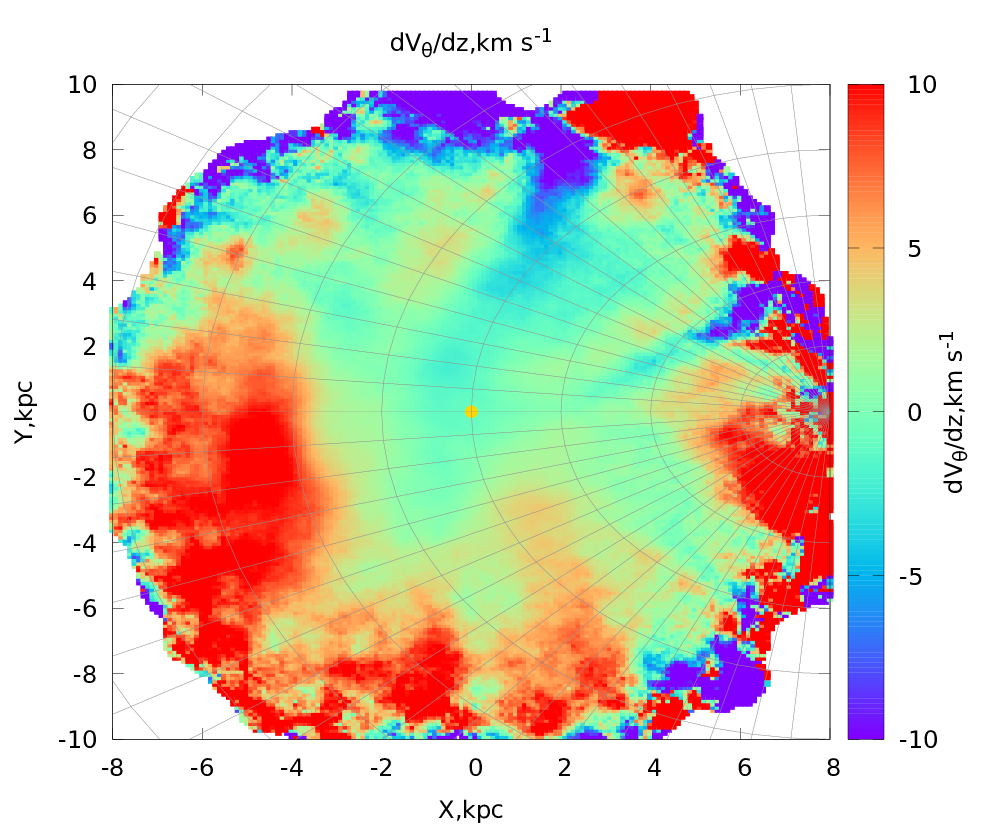}}
  \caption{Vertical gradient of the circular velocity $\frac{\partial V_\theta} {\partial z}$ as a function of the Galactic coordinates, derived from the O--M model parameters (equation \ref{eq:dVtheta_dz_om}, left panel) and from equations \ref{eq:derivatives} (right panel). $R_{\odot}$ is taken to be 8.0~kpc and showed yellow point.}
\label{fig:dVtheta_dz}
\end{figure*}

This gradient is determined from the linear combination of the O--M model parameters $\omega_1$ and $M^+_{23}$ (equation \ref{eq:dVtheta_dz_om}) and from the solution of the system of equations \ref{eq:derivatives}. It is presented in the left and right panels of the Fig. \ref{fig:dVtheta_dz}, respectively.
It should be noted that the values of this parameter were obtained from the data of stars contained in a sphere with a radius of 1 kpc, the center of which is located in the galactic mid-plane. It means that stars both from the southern northern hemispheres get into this sphere. In the previous article \cite{Velichko2020}, we have noted that the gradients for the southern and northern hemispheres have the same module values, but opposite in signs. Therefore, the values of $\frac{\partial V_\theta} {\partial z}$ derived by us in the central part of the region under study are close to zero, since they are values averaged over the stars from the southern and northern hemispheres. The non-zero values of the gradient at the periphery of the region are most likely caused by systematic errors in the distances to stars and errors in stellar astrometric parameters.

\begin{figure}
   \centering
\resizebox{\hsize}{!}
   {\includegraphics{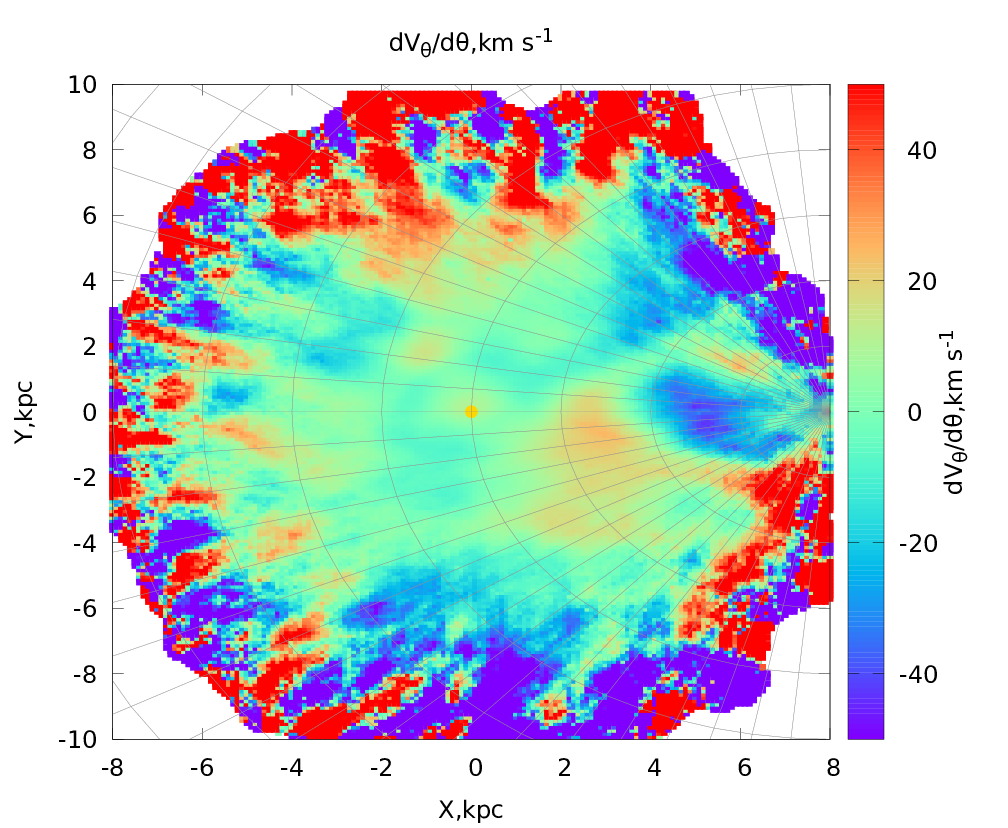}}
  \caption{Azimuth gradient of the circular velocity $\frac{\partial V_\theta} {\partial \theta}$ as a function of the Galactic coordinates derived from equations \ref{eq:derivatives}. $R_{\odot}$ is taken to be 8.0~kpc and showed yellow point.}
\label{fig:dVtheta_dtheta}
\end{figure}

\subsection{Azimuth gradient of the circular velocity component $\partial V_\theta / \partial \theta$}

As mentioned above, the parameter $\frac{\partial V_\theta} {\partial \theta}$ cannot be determined from the O--M model without additional assumptions. The approach implemented by us allows us to define it without any additional conditions. The parameter is shown in Fig. ~\ref{fig:dVtheta_dtheta}. It is clearly seen that the distribution of this parameter does not have obvious regularity. Velocity values in the central part of the area under study vary from -20 to +20 \kms.

\begin{figure}
\resizebox{\hsize}{!}
   {\includegraphics{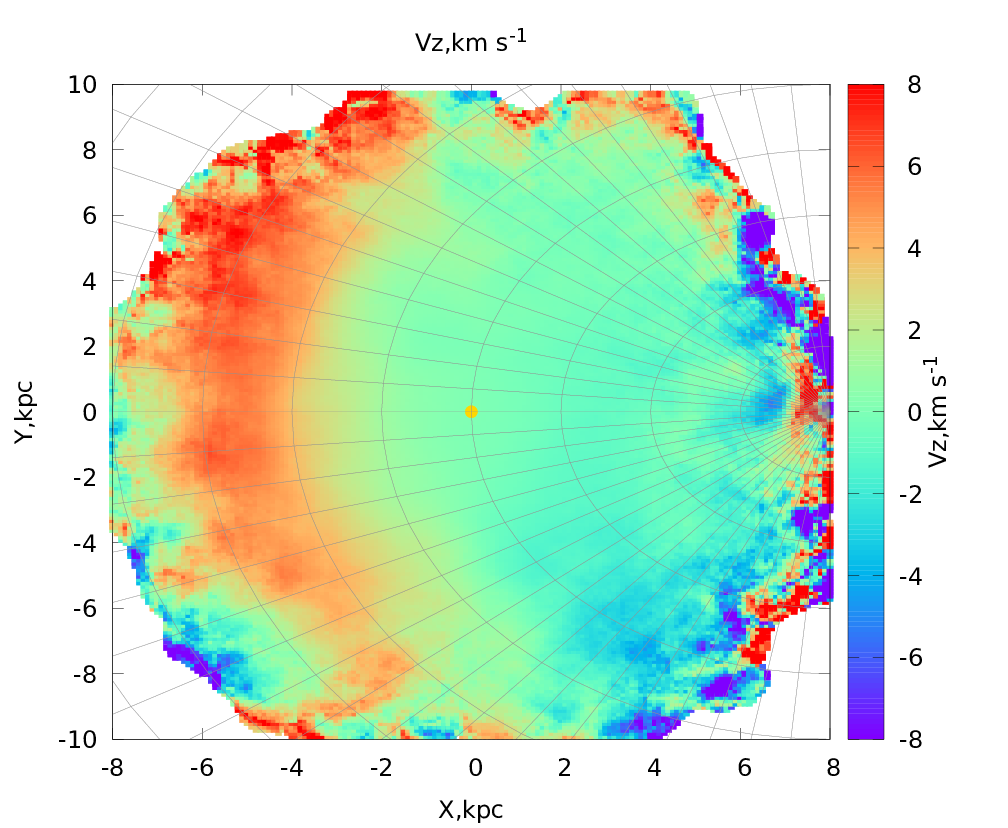}}
  \caption{Vertical components of the centroids velocity as a function of the Galactic coordinates. $R_{\sun}$ is taken to be 8.0~kpc and showed yellow point.}
\label{fig:Vz}
\end{figure}

\begin{figure*}
{\includegraphics [width = 88mm] {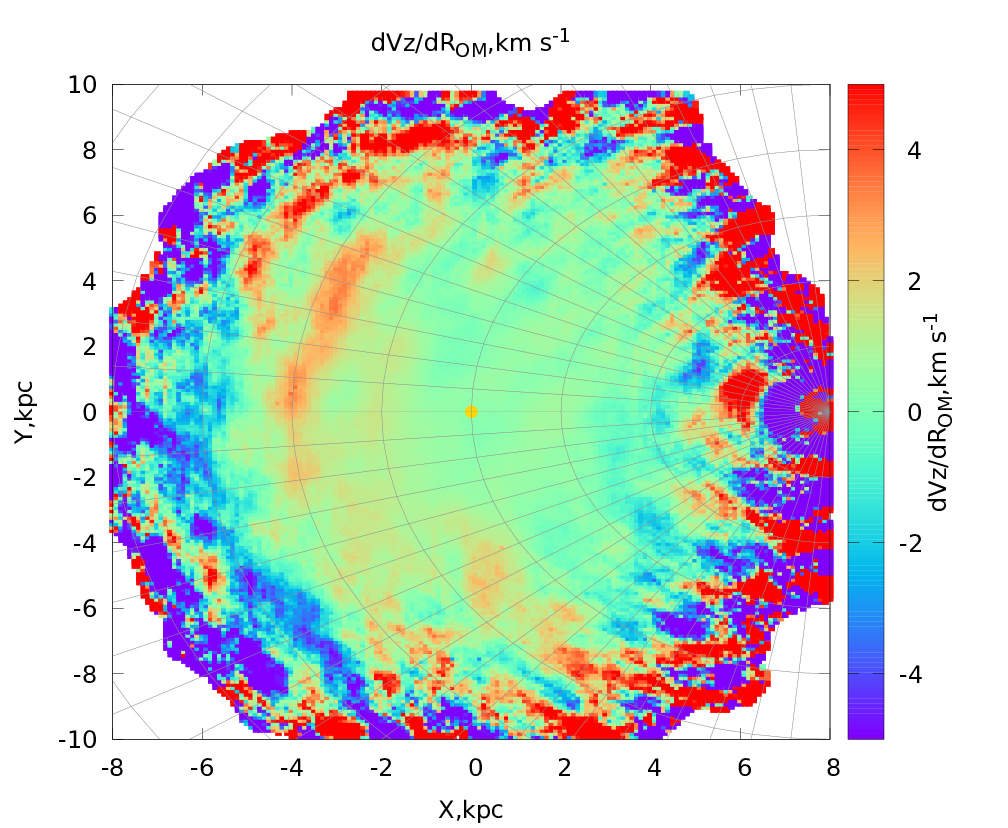}
 \includegraphics [width = 88mm] {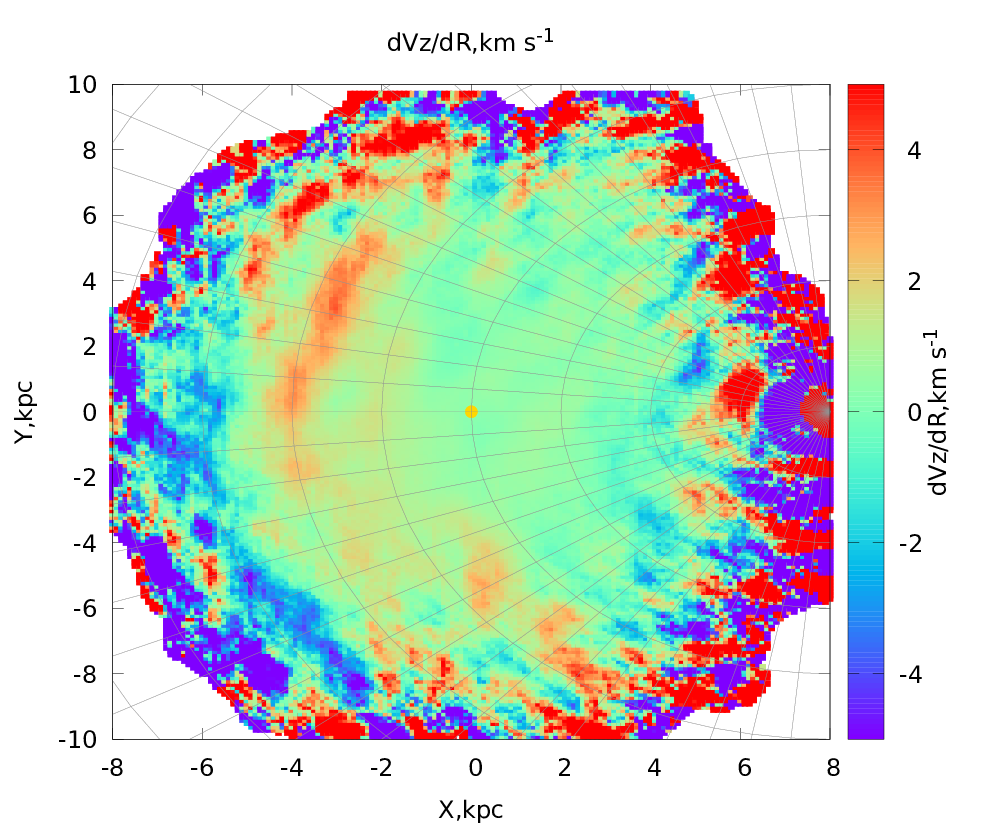}}
\caption{Radial gradient of the vertical centroids velocity component $\frac{\partial V_z} {\partial R}$ as a function of the Galactic coordinates, derived from the O--M model parameters (equation \ref{eq:dVz_dR_om}, left panel), and from the equations \ref{eq:derivatives}. $R_{\odot}$ is taken to be 8.0~kpc and showed yellow point.}
\label{fig:dVz_dR}
\end{figure*}

\subsection{ Vertical component of centroids' velocity $V_z$}

The vertical component of centroids' velocity $V_z$ within the range of Galactocentric distances from 2 to 12 kpc practically does not change (Fig. \ref{fig:Vz}). Outside this range, the velocity values reach $\pm$8 \kms. This is especially well seen beyond 12 kpc, where the increase in velocity is observed as a certain structure located along concentric circles. The value of the vertical velocity at the distance of the centroid of the Sun coincides with the values given in the previous paper \cite{Fedorov2021}, and its behavior in the galactic plane is close to the values obtained in the work by \cite{Poggio2018}.

\subsection{ Radial gradient of the vertical component $\partial V_z / \partial R$}

The gradient of the vertical component of the centroid velocity field along the radius vector is defined from combination of the O--M model parameters $\omega_2$ and $M^+_{13}$ (equation \ref{eq:dVz_dR_om}).

As can be seen from Fig. \ref{fig:dVz_dR}, its values are close to zero in most of the considered galactic plane. And only on the periphery of the area under study the values reach $\pm$4 \kms. Its values obtained by both methods (left and right panels of Fig. \ref{fig:dVz_dR}) practically coincide.

\subsection { Azimuth gradient of the vertical velocity component $\partial V_z / \partial \theta$}

\begin{figure}
   \centering
\resizebox{\hsize}{!}
   {\includegraphics{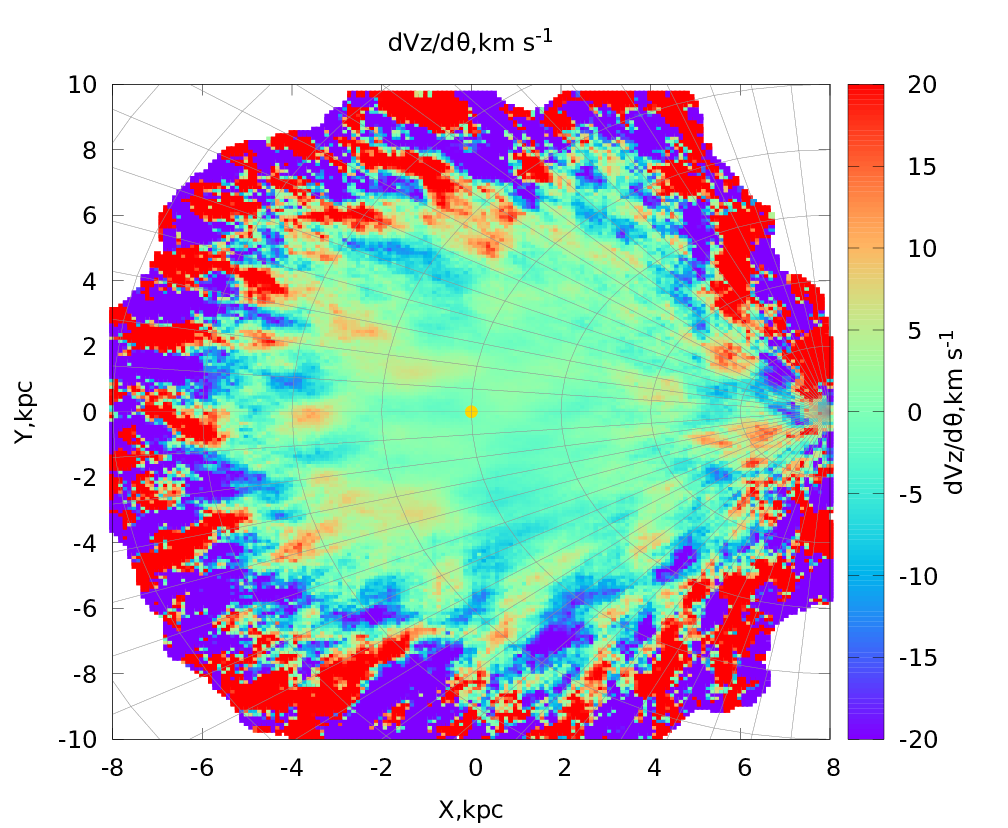}}
  \caption{Azimuth gradient of the vertical centroids velocity component as a function of the Galactic coordinates, derived from the O--M model parameters (equation \ref{warp}, left panel) and from the equations \ref{eq:derivatives} (right panel). $R_{\odot}$ is taken to be 8.0~kpc and showed yellow point.}
\label{fig:dVz_dtheta}
\end{figure}

The product of $R$ and a linear combination of the O--M model parameters from the equation \ref{warp} is a manifestation of the Galactic warp. We note that in the region around the Sun up to heliocentric distances of about 5 kpc, the distribution of the parameter looks relatively smooth, and in the central part of Fig. \ref{fig:dVz_dtheta} only slight deviations of velocities from the background ones are observed. The left and right panels of Fig. \ref{fig:dVz_dtheta} show the parameter distributions obtained by both methods.

\subsection{Vertical gradient of the vertical velocity component $\partial V_z / \partial z$} 

\begin{figure*}
   {\includegraphics [width = 88mm] {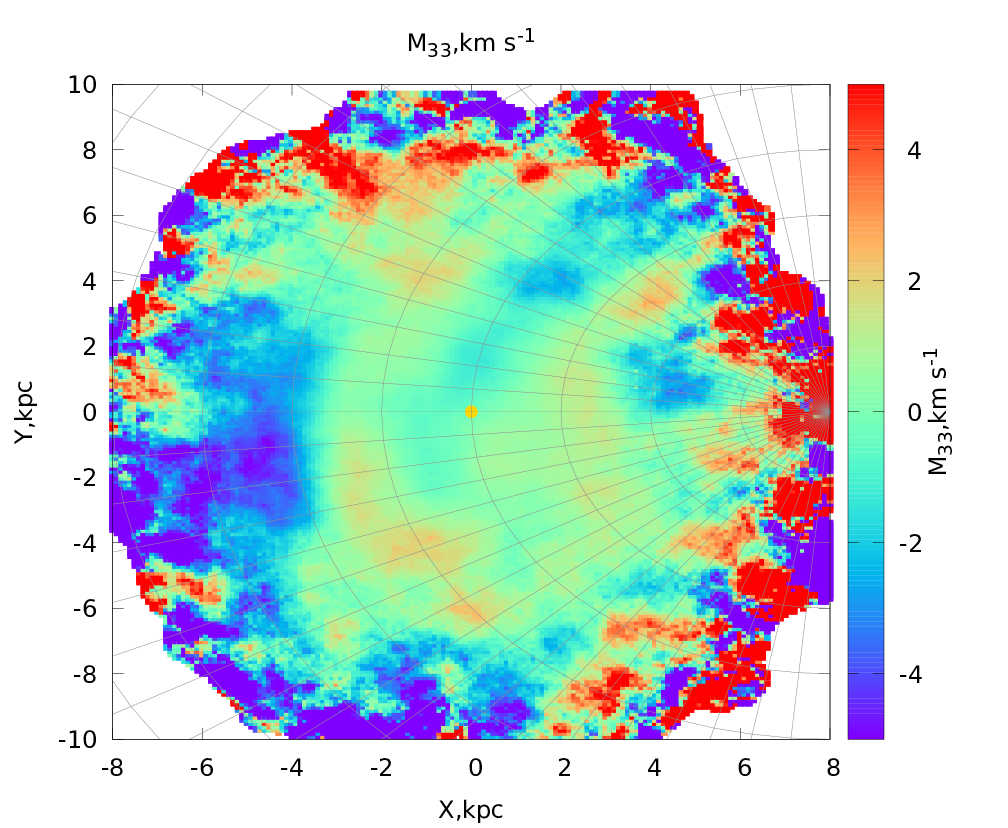}
    \includegraphics [width = 88mm] {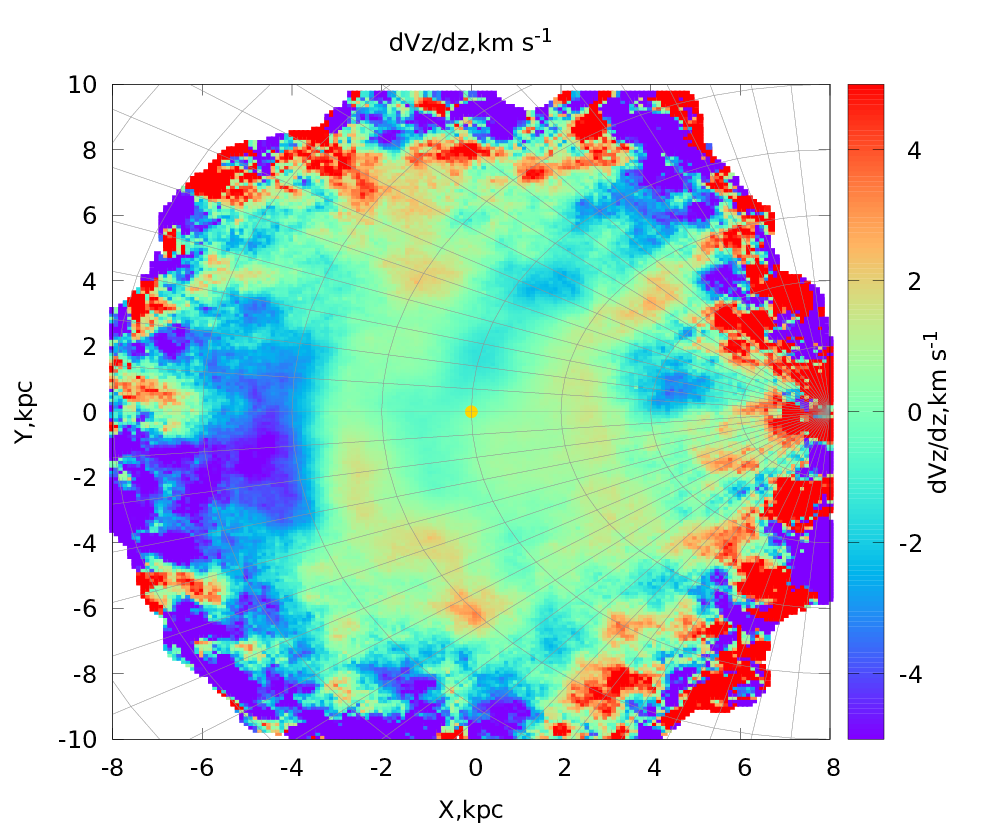}}
  \caption{Vertical gradient of the vertical centroids velocity component as a function of the Galactic coordinates, derived from the O--M model parameters (equation \ref{eq:dVz_dz}, left panel) and from equations \ref{eq:derivatives} (right panel). $R_{\odot}$ is taken to be 8.0~kpc and showed yellow point.}
\label{fig:dVz_dz}
\end{figure*}

The relation \ref{eq:dVz_dz} is interpreted within the O--M model as a compression-expansion of the stellar system along the $z$ axis. As in the previous figures, the panels in Fig. \ref{fig:dVz_dz} show the distributions of the parameter in the galactic plane. Fig. \ref{fig:dVz_dz} shows that the values of this parameter reach $\pm$4 \kms. Some concentric structures with slightly different velocities are also visible.

\section {Summary and conclusions}

In this paper, we performed a kinematic analysis of red giants and subgiants whose centroids are located in the Galactic mid-plane. For the stars of the centroids under study, the spatial coordinates and velocities were taken from the $Gaia$ EDR3 catalogue. We used two methods for determining the kinematic parameters. The first method is the calculation of kinematic parameters within the O--M model, the equations of which are presented in a local rectangular coordinate system. The second method was proposed by us. It is based on computing the average values and gradients of the centroid velocity components in the Galactocentric cylindrical coordinate system. It was shown that the kinematic parameters obtained using these methods are in good agreement with each other. In addition, for the first time, all kinematic parameters as a function of the Galactocentric coordinates were obtained by our method, including those that, in principle, cannot be estimated within the O--M model. These parameters include the gradients of the radial $\frac{\partial V_R} {\partial \theta}$ and the azimuth $\frac{\partial V_\theta} {\partial \theta}$ velocity along the coordinate angle $\theta$. Previously, to determine the rotation velocity of the Galaxy, the value of the parameter $\frac{\partial V_R} {\partial \theta}$ was taken equal to zero. Our approach made it possible to derive the values of these parameters not only within the Solar neighborhood, but also in that part of the Galactic plane, which is provided by 6D astrometric parameters from $Gaia$ EDR3. Non-zero significant values of these parameters indicate the asymmetry of the velocity field in the Galaxy, the complexity of its kinematic characteristics and require further detailed study.
The presented method for analyzing the spatial velocities of stars has made it possible to obtain a number of global characteristics related to much larger volumes of the Galaxy. Also, this approach demonstrated the emergence of new, alternative possibilities, in particular, for determining the behavior of the rotation curve of the Galaxy and its inclination. Estimates of the numerical values of the derived kinematic parameters and their errors can be provided to interested readers by personal request via e-mail: akhmetovvs@gmail.com.
It should be emphasized that in the color figures we show the nature of the behavior of the kinematic parameters. In this case, we are interested exactly in their behavior in the galactic mid-plane. We understand that the module values of the found kinematic parameters depend on the accepted Galactocentric distance $R_\odot$ and the solar velocity components ($V_{x,\odot}, V_{y,\odot}, V_{z,\odot}$). Therefore, the final interpretation of the derived results requires additional studies of the stellar velocity field and the involvement of new data. In the near future, we expect to publish the results of study of some kinematic parameters, which have been given in this work without their deep analysis.

\section{Acknowledgements}
This work has made use of data from the European Space Agency (ESA) mission {\it Gaia} (\url{https://www.cosmos.esa.int/gaia}), processed by the {\it Gaia} Data Processing and Analysis Consortium (DPAC,\url{https://www.cosmos.esa.int/web/gaia/dpac/consortium}). Funding for the DPAC has been provided by national institutions, in particular the institutions participating in the {\it Gaia} Multilateral Agreement.

We are grateful to Ukrainians who are fighting to stop the war so that we can safely finish this article.

%We sincerely thank the anonymous reviewer for a careful reading and insightful comments and suggestions on this paper.

\section*{Data availability}
\addcontentsline{toc}{section}{Data availability}
The used catalogue data is available in a standardised format for readers via the CDS (https://cds.u-strasbg.fr).
The software code used in this paper can be made available upon request by emailing the corresponding author.

\end{document}